\documentclass{IEEEtran}
\usepackage{amsmath}
\usepackage{amssymb}
\usepackage{graphicx}
\usepackage{tabularx}
\usepackage{bm}
\usepackage{xcolor}
\usepackage{cite}
\graphicspath{{./Figures/}}
\usepackage{comment}
\DeclareGraphicsExtensions{.pdf,.jpg,.png,.jpeg}

\begin{document}

\title{
Bayesian estimates of transmission line outage rates that consider line dependencies {\large PREPRINT NOVEMBER 2019}}
\author{Kai Zhou, James R. Cruise, Chris J. Dent, Ian Dobson, Louis Wehenkel, Zhaoyu Wang, Amy L. Wilson\thanks{
KZ,ID,ZW are with Iowa State University, Ames IA USA; dobson@iastate.edu.
JRC is with Riverlane Research, UK.
CJD and ALW are with University of Edinburgh, Scotland. 
LW is with University of Li\`ege, Belgium.

We thank the Isaac Newton Institute for Mathematical Sciences, Cambridge, for support 
during the programme Mathematics of Energy Systems where work on this paper was initiated. This work was supported by EPSRC grant EP/R014604/1.\&\#34.
KZ,ID,ZW 
acknowledge support from USA NSF grants 1609080 and 1735354.
LW acknowledges the support of F.R.S.-FNRS 
and the Simons Foundation.
We gratefully thank Bonneville Power Administration for making publicly available the outage data  that made this paper possible. 
The analysis and conclusions are strictly those of the authors and not of BPA. 
}}
\maketitle

\begin{abstract}
\looseness=-1
Transmission line outage rates are fundamental to power system reliability analysis.
Line outages are infrequent, occurring only about once a year, so outage data are limited. 
We propose a Bayesian hierarchical model that leverages line dependencies to better estimate outage rates of individual transmission lines from limited outage data.
The Bayesian estimates have a lower standard deviation than estimating the outage rates simply by dividing the number of outages by the number of years of data, especially when the number of outages is small.
The Bayesian model produces more accurate individual line outage rates, as well as estimates of the uncertainty of these rates. Better estimates of line outage rates can improve 
system risk assessment, outage prediction, and maintenance scheduling.
\end{abstract}


\section{Introduction}
\label{sec:intro}
Transmission line outage rates are foundational for many reliability calculations, but in historical data the counts of outages for the more reliable lines are low, and estimated individual line outage rates are highly uncertain.
There are several ways in which individual transmission lines are partially similar, including their length, rating, geographical location, and their proximity. We leverage these partial similarities with a Bayesian hierarchical method to improve the estimation of line outage rates from historical data. 

The conventional method of estimating annual line outage rates divides the number of outages by the number of years of data. However, these estimates have a high variance when the data are insufficient. Indeed, in a year, many lines either do not fail or only fail once. 

One pragmatic approach to mitigate the problem of limited outage counts is to group or pool lines together to get an estimate for the outage rate of that group. 
The lines can be grouped by area \cite{ZhouPS06,LiIBM10,DokicHICSS19}, or by line voltage rating.
Lines in the same area experience similar weather conditions, and lines of the same rating have similar construction.
However, the similarity between lines in these groups is only partial, variations of outage rates within the groups are neglected, and it is unwieldy to group lines according to multiple characteristics.

Transmission line outage rates are often supposed to be proportional to line length, and they are often quoted as rates per unit length \cite{IesmantasEPSR19}. However, a line's outage rate is not strictly proportional to the line length because of substation and other effects, making the dependence on line length only a partial dependence. 
Indeed, our historical line outage data shows only a limited dependence on line length.

\looseness=-1
There is a middle ground between pooling lines in groups assuming perfect line dependencies within the group, and completely neglecting dependencies between lines by computing individual line outage rates in isolation. 
To exploit the partial dependencies of line outage rates,
this paper proposes a Bayesian hierarchical method to estimate outage rates of individual transmission lines. 
In particular, our method can leverage the multiple partial dependencies in line length, rating, network proximity, and geographical area to give better outage rates of individual lines. This is done by explicitly modeling the dependence of outage rates on line length and rating and by using covariance kernels to model the dependencies between lines in close proximity. Our method can, therefore, learn about the outage rates of individual lines from lines close-by and with similar lengths and ratings. This means that where there is little data associated with a line (because the outage rate is small), our method can still estimate an outage rate for that line and its uncertainty. Also, by borrowing information from other lines, we can expect smaller uncertainties associated with estimates of outage rates, without assuming that all lines within a group have the same outage rate (as would be the case if we pooled the data). 

After reviewing the literature in Section \ref{sec:literature}, Section \ref{sec:data} presents the historical outage data collected by a large utility and the modeling of the line dependencies.
Sections \ref{sec:model} and \ref{sec:processing} present the Bayesian hierarchical model and the processing of the utility data. 
As we do not know the true outage rates from historical outage data, we use synthetic data to validate and evaluate the performance of the Bayesian hierarchical model in Section \ref{sec:synthetic}. Section \ref{sec:conclusion}  concludes the paper.

\section{Literature review}
\label{sec:literature}

\looseness=-1
Bayesian approaches encode uncertainty in uncertain parameters such as outage rates as random variables. The Bayesian analysis aims to estimate a probability distribution for the uncertain parameters by incorporating all of our knowledge and accurately reflecting the uncertainty. 
Bayes theorem is used to combine data with prior distributions that describe initial knowledge of the uncertainty. The prior distributions are updated with the available data to give a posterior distribution that describes the uncertainty in the parameter values given all the available data.
The mean or mode of the posterior distribution can be used to give a point estimate of the parameter.
For further detail explaining Bayesian methods we suggest \cite{carlin08BMDA} as an introduction and \cite{GelmanBDA13} as a reference.

\begin{table*}[t]
	\centering
	\caption{Annual Outage Counts, Line attributes, and Bayesian estimates of outage rates after 1st, 7th and 14th years for 4 lines}
	\begin{tabular}{ccccccccccccccccccccc}
		Line & \multicolumn{14}{c}{Outage counts in different years}                   & \multicolumn{3}{c}{Line attributes} & \multicolumn{3}{c}{Annual outage rate} \\ 
		ID & 1 & 2 & 3 & 4 & 5 & 6 & 7 & 8 & 9 & 10 & 11 & 12 & 13 & 14 & Voltage(kV)\!\!\!& \!\!\!Length(mile)\!\!\!& \!\!\!District& 1st & after 7th  & after 14th \\ \hline
		29  & 0 & 0 & 0 & 0 & 0 & 0 & 0 & 0 & 3 & 2 & 0 & 0 & 0 & 0 & 230 & 8.3 & P& 0.32 & 0.17 & 0.37 \\
		11  & 0 & 0 & 1 & 0 & 0 & 1 & 0 & 0 & 1 & 0 & 0 & 0 & 0 & 1 &500 & 22.65& N& 0.36 & 0.33 & 0.34 \\
		2 & 1 & 2 & 0 & 0 & 0 & 0 & 0 & 0 & 0 & 0 & 0 & 0 & 0 & 0  &230 &7.62 & A& 0.73 & 0.48 & 0.28 \\
		8 & 1 & 2 & 4 & 2 & 1 & 2 & 2 & 2 & 2 & 1 & 3 & 8 & 6 & 2 &500 & 148.86& E& 0.93 & 1.85 & 2.56 \\ \hline
	\end{tabular}
	\label{tbl:bayesest}
\end{table*}
\looseness=-1
Bayesian methods are ideal for problems with limited data (such as estimation of outage rates), where it is necessary to use all the information available. 
Studies in ecology and social science have shown that when data are limited, Bayesian methods have less bias and are more robust than frequentist methods that consider parameters as fixed values \cite{OmlinEM99,StegmuellerAJPS13}. When lots of data are available, the data outweighs any effect of the prior distributions and a Bayesian method is less advantageous. 
 
There is previous research predicting outage rates 
using Bayesian methods. 
Zhou \cite{ZhouPS06} proposes a simple Bayesian network to predict weather-related outage rates given lightning and wind conditions over the whole system. 
It compares the Bayesian network with a Poisson regression model and concludes that the Bayesian network is preferable. 
Li \cite{LiIBM10} and Ie{\v{s}}mantis \cite{IesmantasEPSR19} present two Bayesian hierarchical models. \cite{LiIBM10} develops a hierarchical model to predict outage counts in a district given weather conditions. Our model, however, inspects individual transmission lines by leveraging the correlation between these lines, which is more granular than modeling the correlation between areas. 
Ie{\v{s}}mantis \cite{IesmantasEPSR19} presents a Poisson-gamma random field model to estimate 230~kV transmission line outage rates per kilometer. In contrast, our paper uses lines with all the high voltage ratings and does not assume that outage rates are proportional to line length. 

\looseness=-1
Transmission line outages are correlated with each other in several ways. Lines in the power grid interconnect at substations, and some faults or substation arrangements may trip several lines simultaneously. Multiple line outages also occur due to protection schemes such as control protection groups and remedial action schemes.
Moreover, lines in the same area experience similar weather conditions. There is some previous work on these dependencies. Li \cite{LiIBM10} uses the network adjacency matrix to model district dependencies.
Similarly, Dokic \cite{DokicHICSS19} uses the weighted adjacency matrix to model substation dependencies. The difference between them is that \cite{LiIBM10} models the dependencies as a covariance matrix from the Bayesian perspective, while \cite{DokicHICSS19} uses an embedding method by learning vector representations of dependencies from a frequentist perspective. 
But both \cite{LiIBM10} and \cite{DokicHICSS19} study pooled outage rates for different areas.   
Ie{\v{s}}mantas \cite{IesmantasEPSR19} models geographical dependencies between the outage rate per kilometer of 230~kV lines by making a rectangular grid of the area.
Portions of lines in the same rectangle are assumed to have the same geographical influence, and the correlation between lines in different rectangles is assumed. 
The main conclusion of \cite{IesmantasEPSR19} is that geographical correlation between line outage rates is present but weak.

\section{Exploring historical outage data and modeling line dependencies}
\label{sec:data}

Utilities routinely collect detailed outage data. For example, NERC's Transmission Availability Data System (TADS) collects outage data from North American utilities.
Here, to illustrate our methods, we use some publicly available historical line outage data \cite{bpadata}.

\subsection{Historical outage data}
The historical line outage data we use consists of transmission line outages recorded by a North American utility \cite{bpadata} for fourteen years since 1999.  The data record forced and scheduled line outages, including the sending and receiving bus names of outaged lines, outage start and end times and dates,  line attributes such as lengths, voltage ratings, districts in which a line is, and outage causes. Some lines cross several districts.
There are 549 lines outaging in the data with rated voltages of 69, 115, 230, 287, 345, and 500~kV.

We neglect the scheduled outages and only consider the forced line outages. We also exclude the two 1000~kV HVDC lines, and momentary outages (outage duration does not exceed one minute). 
There are lines that failed once or twice in most of the years but suddenly failed, for example, ten times in one year. 
One common reason that a line could fail several times in a day is outages and  reclosures for the same cause.
So if a line fails several times in a day, we only count it once. 
Table \ref{tbl:bayesest} shows an example of the outage data.

\subsection{Data exploration}
\label{sec:dataexplore}
We initially explore the line outage data using the conventional method of estimating annual line outage rates by dividing the number of outages by the number of years of data.
We first pool all the line data together (i.e. treat as one homogeneous data set) to calculate the overall mean and standard deviation of outage rates, which are 0.6 and 0.7 outages per year, respectively. 
Next, we examine the individual conventional line outage rates.
The mean variance-to-mean ratio of outage counts for each line is 1.2, which indicates that the outage counts show some overdispersion\footnote{Overdispersion means that the variance is larger than the mean. The Poisson distribution commonly used for count data does not apply when there is overdispersion because the Poisson mean and variance are equal.}.
%

\begin{figure}[htb]
	\centering
	\includegraphics[width=\columnwidth]{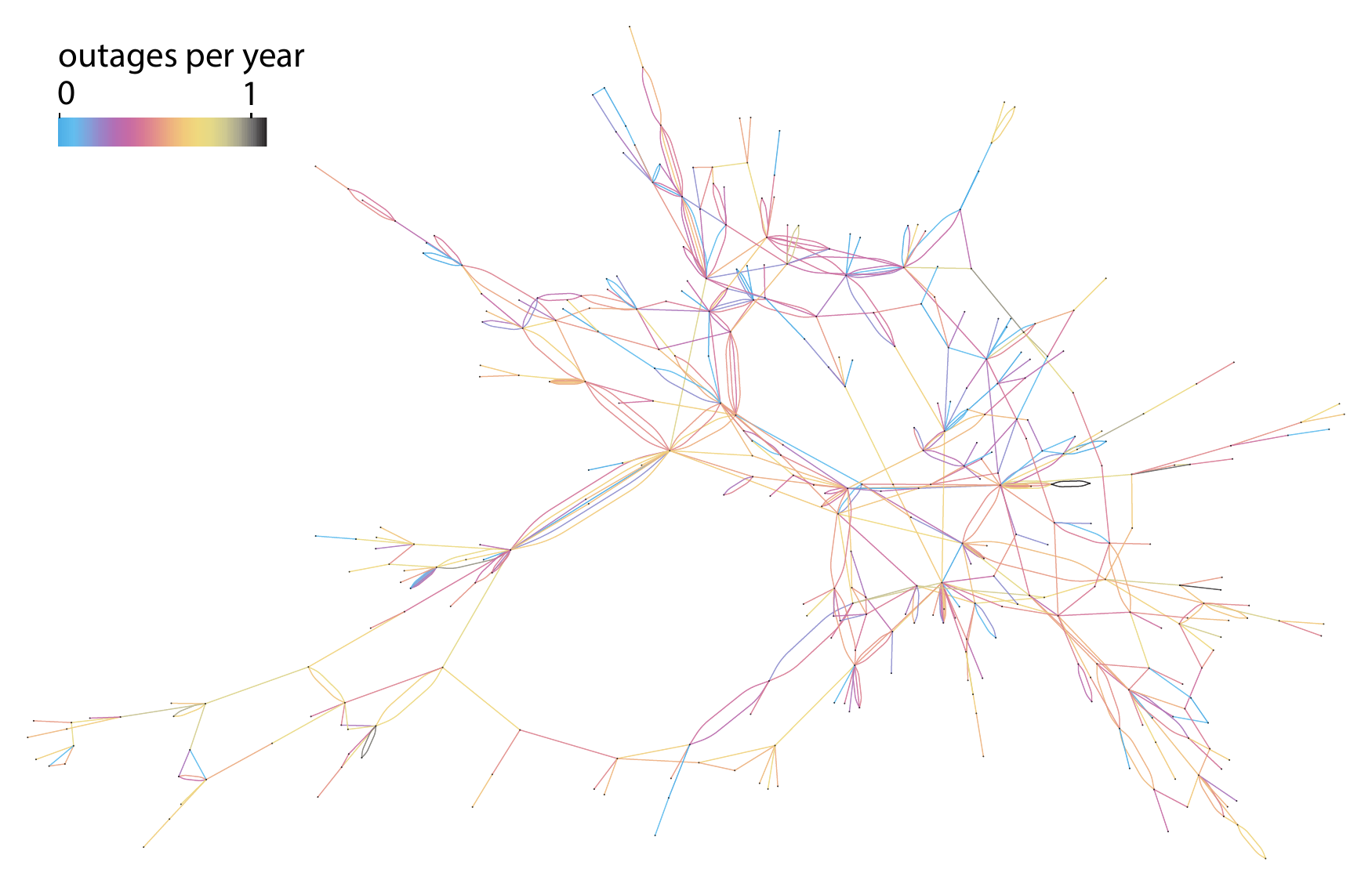}
	\caption{Annual outage counts indicated by different colors (Network layout is not geographic.)}
	\label{fig:visRates}
\end{figure}

\looseness=-1
The power system network can be deduced directly from the outage data using the method of \cite{DobsonPS16}, and we show the conventional outage rates on the network in Figure \ref{fig:visRates} to visualize the spatial correlation.
Close lines tend to have close colors, which indicates line dependencies from network proximity.

\subsection{Scaling line lengths and voltage ratings}
\label{sec:scaling}
The line lengths and voltage ratings are transformed and scaled so that their magnitudes and variations are scale-free and comparable. We do this in ways suggested by Gelman \cite{gelman19prior} for generic priors.  

Line lengths in the vector $\bm L$ are first transformed by the natural logarithm to make the range of values less  extreme, and 
then divided by the scale so that their variations are order of magnitude one: 
\begin{align}
\bm{x}_L= \frac{\ln{\bm L}}{{\rm scale}(\ln{\bm L})}
\end{align}
Here the scale of the sample in a vector $\bm z$ is estimated by the Mean Absolute Deviation, which is ${\rm scale}(\bm z) = {\rm median}(\bm z - {\rm median}(\bm z))$.
Note that we use bold variables for vectors in this paper, and functions such as $\ln$ are applied element-wise so that  $\ln{\bm L}= [\ln L_1,...,\ln L_N]'$.

Similarly, the line voltage ratings $\bm V$ are first scaled by ${\rm SD}(\bm V)$, the standard deviation of $\bm V$, and then divided by the scale:
\begin{align}
\bm{x}_V = \frac{{\bm V}/{\rm SD}(\bm V)}{{\rm scale}({\bm V}/{\rm SD}(\bm V))}
\end{align}

It is usually considered that the line length and voltage rating have a positive correlation. Indeed, the BPA data shows this correlation, but it is a weak correlation: the Pearson correlation coefficient is 0.34 (0.12 for transformed lengths and voltage ratings). 

\subsection{Line proximity}
\label{sec:kernels}
The proximity of lines is quantified by the weighted sum of two kernels, which reflect two aspects of proximity. The first kernel is based on districts. Lines in the same district are more likely to experience the same weather conditions. Another kernel is based on network distance in terms of line length, which, to some extent, reflects both geographic proximity and the physical and engineering interactions in the power grid.  We fit a linear regression model with correlated lines (described below) to support the form of the Bayesian hierarchical model and give guidance on setting priors.

\subsubsection{Districts}
There are 12 districts, and districts for each line are represented by a feature vector $\phi_{dis} \in \{0,1\}^{12}$  whose coordinates correspond to the districts, and are set to 1 for each district crossed by that line, and to 0 otherwise. The scalar product in this feature space thus counts the number of common districts crossed by two lines. 

We define the district kernel as:
\begin{align}
    \bm \Sigma_1 = \exp\left(-||\phi_{dis}(i) -\phi_{dis}(j)||_2^2 - \mathbb{I}_{i\neq j} \right)
\end{align}
where $||\cdot||_2$ stands for the two-norm, and $\mathbb{I}_{i\neq j}$ is an indicator function. The reason why $\mathbb{I}_{i\neq j}$ is included is that a line is most similar to itself.
The kernel
$\bm \Sigma_1$ has the form of a correlation matrix since it is positive definite.

\subsubsection{Network distance}
The network distance between lines $L_i$ and $ L_j $ along the network lines is defined as
\begin{align}
d_{ij}=d(L_i,L_j)=&\mbox {minimum length in miles of a network path }\notag\\[-3pt]
&\mbox{ joining  midpoint of } L_i  
\mbox{ to midpoint of } L_j.\notag
\end{align}
For example, the distance  of line to itself is zero and the distance of a line to a neighboring line with at least one bus in  common is half of the total length of the two lines.

Then we use the exponential kernel  $\bm \Sigma_2$ which is 
\begin{align}
    \bm \Sigma_2 = \exp[-2d(L_i,L_j)]
\end{align}
As $d(L_i,L_i)=0$, the diagonal elements of $\bm \Sigma_2$ are one. 

\subsubsection{Combining the two kernels}
The network proximity $\bm \Sigma$ is the weighted  sum of above two kernels:
\begin{align}
\bm \Sigma = w \bm \Sigma_1 + (1-w) \bm \Sigma_2,
\end{align}
where $0 < w < 1$. For example, if the two kernels are equally important, then $w=0.5$.

We find the weights by fitting a linear regression model for the logarithm of average outage counts with $\bm \beta_0$ following a multivariate normal distribution to model correlation:
\begin{align}
\ln{\frac{\bm N}{t}} & = \bm \beta_0 +  \beta_L \bm x_L + \beta_V \bm x_V,\\
\bm \beta_0 &\sim \mathcal{N}(m \mathbf{1}, \sigma^2 \bm \Sigma),
\label{beta0}
\end{align}
where $\bm N$ is a column vector whose entries are the total number of counts in $t$ years, $\mathbf{1}$ is a column vector of ones, $m, \beta_L, \beta_V$ are scalars, 
and
\begin{align}
 \sigma^2 \bm \Sigma= \sigma^2 (w \bm \Sigma_1 + (1-w) \bm \Sigma_2))= \sigma_1^2 \bm \Sigma_1 + \sigma_2^2 \bm \Sigma_2 .
 \label{sigmadecomp}
\end{align}

We decouple the dependencies between different lines in (\ref{sigmadecomp}) by a coordinate transformation to diagonalize the covariance matrix $\sigma^2 \bm \Sigma$. This transforms the multivariate normal random vector $\bm \beta_0$ in (\ref{beta0}) into independent univariate normal random variables in the vector $\bm \beta'_0$. This decoupling facilitates the maximum likelihood calculation below.
In particular,
by simultaneous diagonalization \cite[p.286]{HornMA02}, 
we find a matrix $\bm Q$ such that $\bm Q^T \bm \Sigma_1 \bm Q = \bm I$ and  $\bm Q^T \bm \Sigma_2 \bm Q = \bm \Lambda$, where $\bm \Lambda$ is a diagonal matrix.
Define $\bm \beta'_0 = \bm Q^T \bm \beta_0$, then
\begin{align}
\bm \beta'_0 &\sim \mathcal{N}(m \bm{Q}^T \mathbf{1}, \bm{Q}^T (\sigma_1^2 \bm \Sigma_1 + \sigma_2^2 \bm \Sigma_2) \bm{Q})\notag\\
 &\sim \mathcal{N}(m \bm{Q}^T \mathbf{1},\sigma_1^2 \bm I + \sigma_2^2 \bm \Lambda).
\end{align}
Then we use  maximum likelihood to estimate the parameters $\sigma_1^2, \sigma_2^2$. 
Using the utility data, we have $\sigma_1^2 = 0.45, \sigma_2^2 = 0.42$. By normalizing $\sigma_1^2$ and $\sigma_2^2$, we have $w = 0.52$.
Moreover, $m = -1.5$ and $ (\beta_L, \beta_V) = (0.13, 0.12)$. 
The positive values of $\beta_L$ and $\beta_V$ indicate that longer lines or higher voltage lines tend to have higher outage rates, which is reasonable. 
These values shall give guidance on setting priors in Section \ref{sec:model}.

\begin{figure}[!ht]
\centering
\includegraphics[width=\columnwidth]{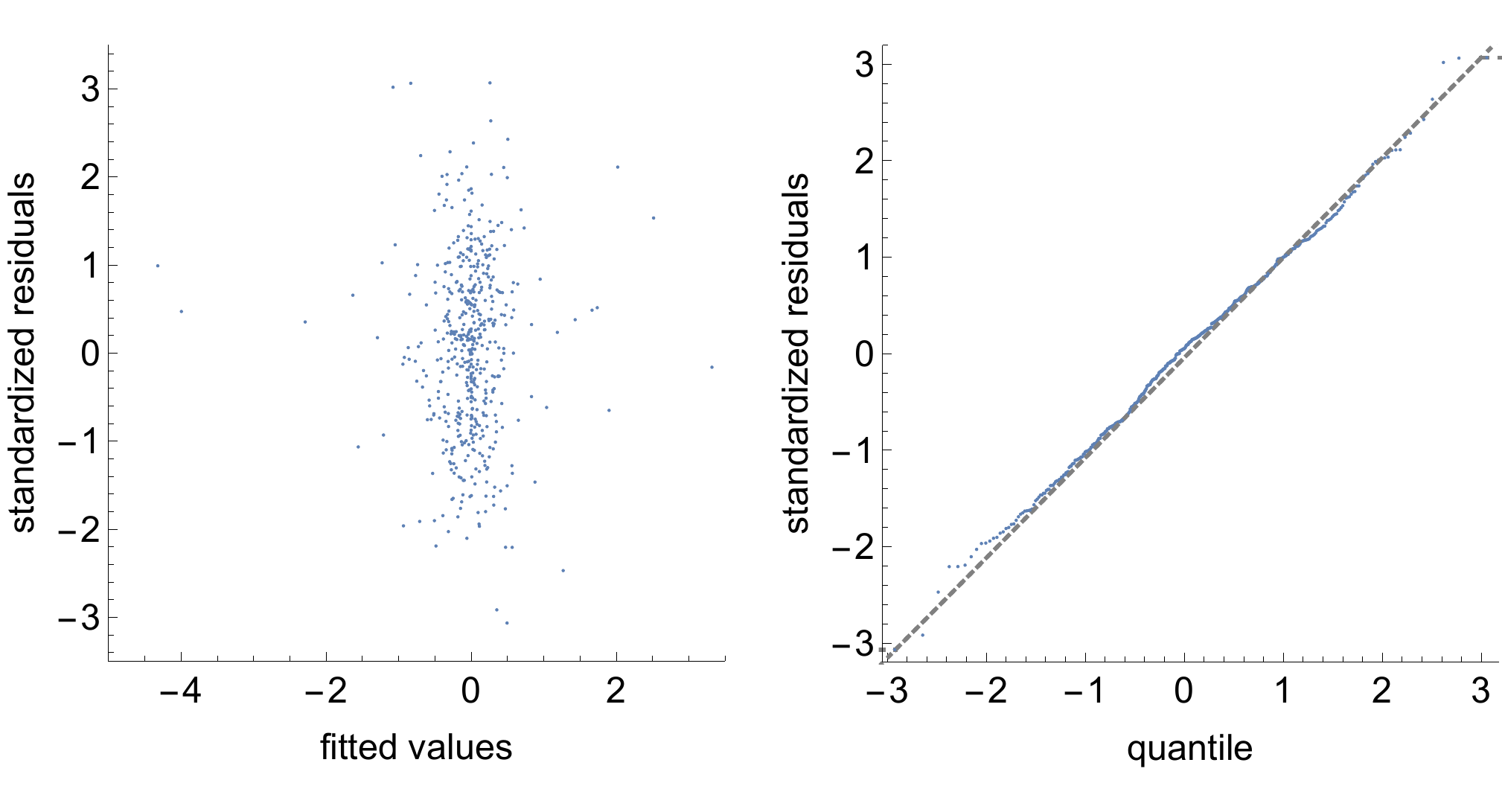}
\caption{Residual plot (left) and QQ-plot (right) for Pearson residuals.}
\label{fig:residual}
\end{figure}

We check the model assumptions by using the residual plot and QQ-plot as shown in Figure \ref{fig:residual}. $\bm \beta'_0$ has no correlation, so we focus on the transformed linear model, and Pearson residuals are used here as $\bm \beta'_0$ has heterogeneous variance. 
The Pearson residual 
is estimated by 
   $ \epsilon'_i =
   \epsilon_i/\sqrt{\sigma_1^2 + \sigma_2^2 \Lambda_i}$,
where the raw residuals are $\bm \epsilon = \bm{Q}^T \ln{\bm N /t} - \bm{Q}^T \bm{X}\bm{\beta} - m \mathbf{1}$ and $\Lambda_i$ is the $i$th diagonal entry of matrix $\bm \Lambda$.
There is no noticeable trend in the residual plot, and the QQ-plot shows that the Pearson residual follows the normal distribution.

\section{The Bayesian hierarchical model with line dependencies}
\label{sec:model}

We propose a Bayesian hierarchical model of outage counts incorporating line dependencies. 

We assume that outage counts follow a Poisson distribution:
\begin{align}
    N_i &\sim {\rm Poisson}(\lambda_i t_i),\quad i=1,...,n,
\end{align}
where $N_i$ is the outage count for line $i$ over $t_i$ years, $\lambda_i$ is the annual outage rate, and $n$ is the number of lines. 

We assume that the outage rates $\lambda_i$ follow a Gamma distribution:
\begin{align}
   \lambda_i &\sim {\rm Gamma}(\alpha, \alpha / \mu_i) ,\quad i=1,...,n
\end{align}
The Gamma distribution is chosen for two reasons:  It is a conjugate prior for the Poisson distribution. Moreover, the Gamma distribution  mean is $\mu_i$ and its variance is $\mu_i^2/\alpha$. The variance of the Gamma distribution increases quadratically as the mean increases, which allows for the overdispersion observed in Section \ref{sec:dataexplore}.

The mean outage rate $\mu_i$ is modeled via a linear regression model with correlated lines.
The linear regression model assumes the predicted variable is normally distributed, but $\mu_i$ is positive and may have a large range of values, so $\mu_i$ is transformed by a log function \cite[Sec.~3.6]{kutner05LM}:
\begin{align}
    \ln\bm \mu & =  \bm \beta_0 + \beta_L \bm x_L  + \beta_V \bm x_V, \label{equ:bayesmodel1}
\end{align}
where $\bm \mu $,
$\bm \beta_0$ are column vectors.

$\bm \beta_0$ follows a multivariate normal distribution: 
\begin{align}
    \bm \beta_0 &\sim \mathcal{N}(m \mathbf{1},\sigma^2 ( w \bm \Sigma_1 + (1-w) \bm \Sigma_2)) \label{equ:bayesmodel2}
\end{align}
Line dependencies are captured by the covariance matrix of this multivariate normal distribution, $\sigma^2$ is a scalar which controls the magnitude of the covariance and $w$ controls the weights of the two kernels. The parameters $\alpha$, $\beta_V$, $\beta_V$, $m$, $\sigma^2$ and $w$ will be estimated using prior distributions in combination with the data as described below. 

The prior distributions are:
\begin{equation}
\begin{aligned}[c]
    \alpha & \sim{\rm Half~Normal}(0.7,8^2) \\
    \beta_L &\sim {\rm Normal}(0.13,5^2)\\
    \beta_V &\sim {\rm Normal}(0.12,5^2)\\
  \end{aligned}
  \quad
\begin{aligned}[c]
        m &\sim {\rm Normal}(-1.5,5^2) \\
    \sigma^2 &\sim {\rm Half~Normal}(0,0.5^2)\\
    w & \sim {\rm Beta}(1,1)
\end{aligned}\notag
\end{equation}
These priors are chosen to ensure that the parameters have a reasonable range and/or mean when compared to our knowledge about the system and the model tested in Section \ref{sec:dataexplore}. As there is not much information about the standard deviations about these priors, we make these priors weakly informative. 
The detail is as follows.

The prior for $\alpha$ is a half-normal distribution with $\alpha>0$. As discussed in Section \ref{sec:dataexplore}, the mean annual outage rate is 0.6, and the standard deviation is 0.7. 
This suggests the expected value of $\mu$ is 0.6, so the expected value of $\alpha$ would be $0.6^2/0.7^2=0.7$ (as $\mu_i^2/ \alpha = {\rm Var}\lambda_i$). 
The standard deviation of $\alpha$ is $\frac{(0.6 + 2 \times 0.7)^2}{0.7^2} - 0.6 \approx 8$ (the numerator is the maximum of $\mu$ in a typical range estimated by two times the standard deviation, $\frac{(0.6 + 2 \times 0.7)^2}{0.7^2}$ is the maximum of $\alpha$). 

Priors for $m, \beta_L, \beta_V$ are normal distributions. The linear regression model in Section \ref{sec:kernels} suggests expected values for these parameters. Considering values of $\bm x_L$, $\bm x_V$, and $\ln{\bm N / t}$ are in $[-10,10]$, we set the standard deviations of $m, \beta_L, \beta_V$ to 5 so that 95\% of the values lie in $[-10,10]$ and they vary mostly in the same magnitude, which produces weakly informative priors. 

$\sigma^2$ functions as a variance. The inverse-gamma prior is usually preferred since it is a conditional conjugate distribution. Gelman \cite{GelmanBA06}, however, does not recommend the inverse-gamma prior as the estimation of $\sigma^2$ would be sensitive to the parameters of inverse-gamma distribution when $\sigma^2$ is near zero. 
Thus, we let $\sigma^2$ have a half-normal prior. Section \ref{sec:kernels} shows that $\sigma_1^2$, $\sigma_2^2$ are about 0.5, so we set the standard deviation of $\sigma^2$ to 0.5 to make at least 95\% of the values of $\sigma^2$ to lie in $[0,1]$.  

We give $w$  a uniform prior as we know that $w$ lies in $[0,1]$ and the expectation of $w$ is $0.52\approx0.5$ from Section \ref{sec:kernels}. 

\section{Bayesian Processing of real data}
\label{sec:processing}

The Bayesian hierarchical model described in the previous section is applied to the historical outage data. We use Monte Carlo Markov Chain (MCMC) to sample from the posterior distribution \cite{GelmanBDA13}. MCMC is a class of algorithms for sampling from a probability distribution. It is used to calculate high-dimensional integrals, which makes it possible to draw samples from the posterior distributions of large Bayesian hierarchical models. This model is implemented in MathematicaStan \cite{stan}, which uses Hamiltonian Monte Carlo as its MCMC algorithm.

\looseness=-1
We sample 2000 times, and the first 1000 samples are burn-in. Appendix \ref{app2} discusses technical details of model diagnostics and algorithm convergence. In this section, we focus on the result of the sampling.

\begin{figure}[!ht]
\centering
\includegraphics[width=\columnwidth]{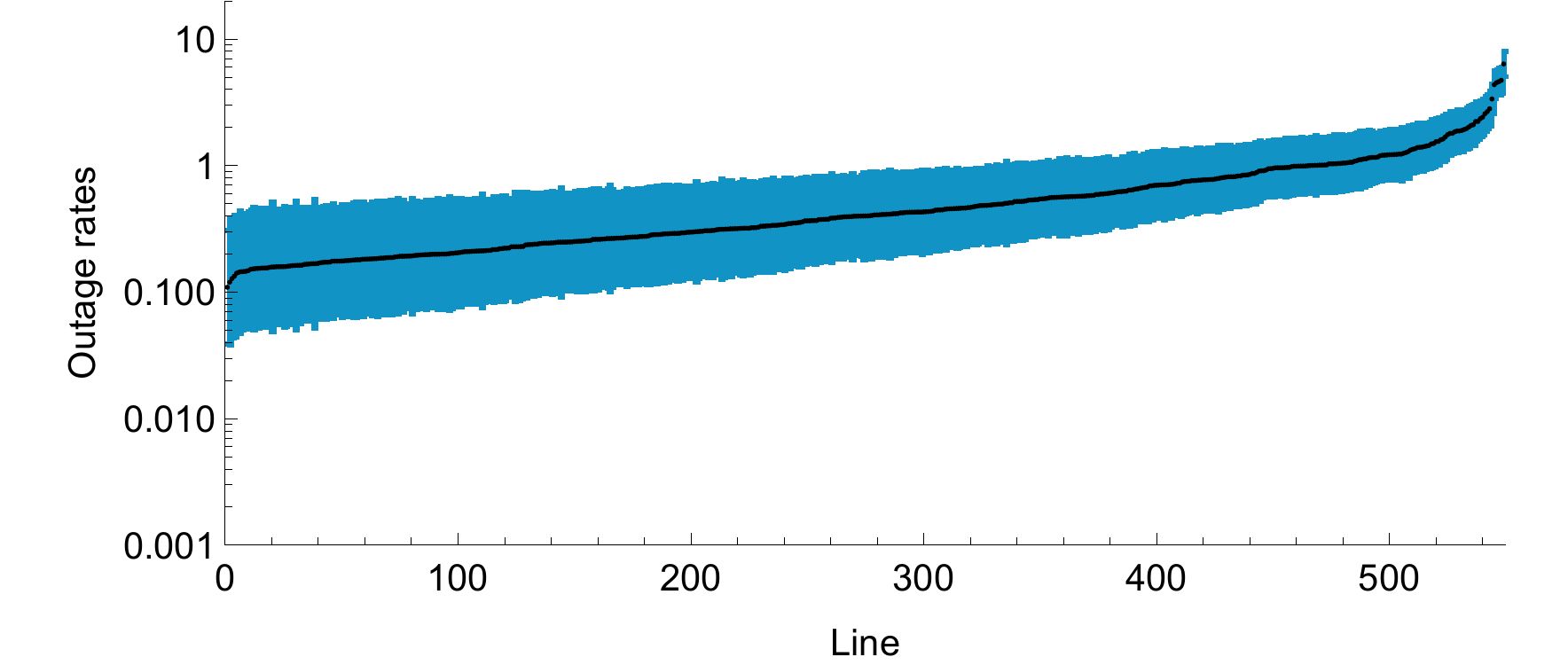}
\caption{Outage rate point estimates (black dots) and 95\% credible intervals (blue bars). Lines are ordered by point estimates.}
\label{fig:cibpa}
\end{figure}

\looseness=-1
We use the posterior mean as the point estimate of line outage rate because 
the posterior mean minimizes the Bayes risk in terms of squared error loss.
Figure \ref{fig:cibpa} shows the point estimates of line outage rates and their 95\% credible intervals\footnote{The credible interval is described by  the multiplicative factor $\kappa$ within which the outage rate ${\lambda}_i$ can vary from the point estimate $\hat\lambda_i$ with 95\% probability; that is, 
$P[\hat\lambda_i/\kappa \leq {\lambda}_i \leq \hat\lambda_i \kappa] = 95\%$. 
}.
The mean outage rate of all lines is 0.74 outages per year, and 82\% of lines have rates less than 1 outage per year. 
There are two lines with very high outage rates. By inspecting the cause codes of these outages, one line outaged mainly because of foreign trouble (which is an external cause outside the power system, such as vehicles striking towers), while the other outaged mainly because of a remedial action scheme.

\begin{figure}[!ht]
\centering
\includegraphics[width=\columnwidth]{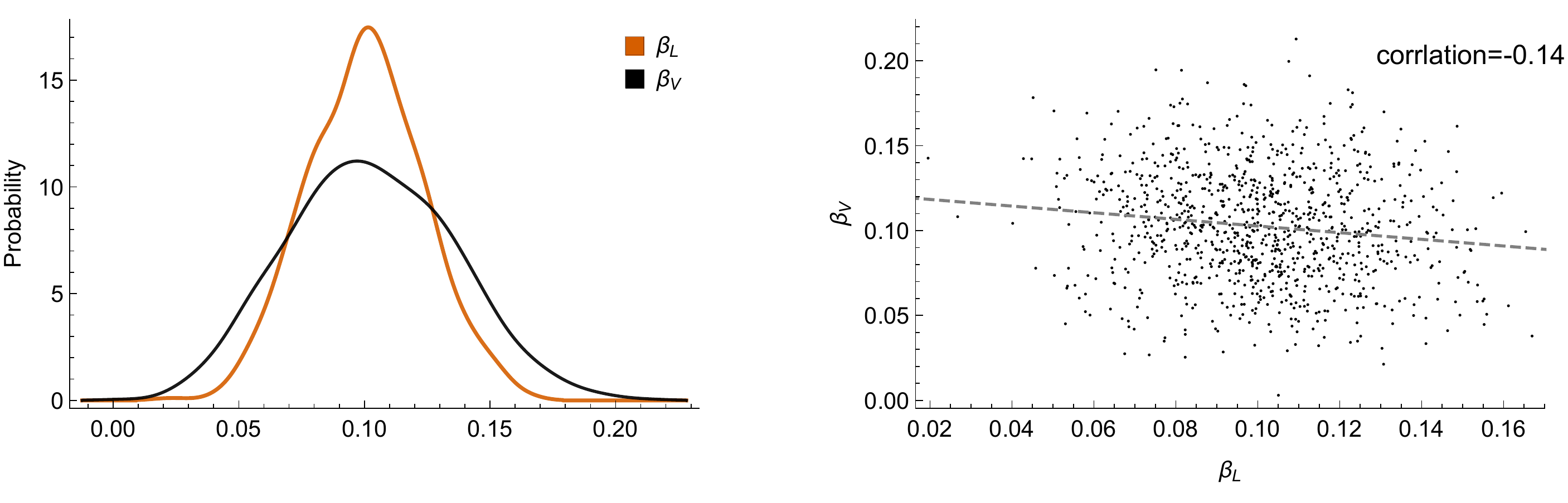}
\caption{Distributions of $\beta_L$ and $\beta_V$ (left) and their scatter plot and  correlation (right). }
\label{fig:pdfbetas}
\end{figure}

The values of $\beta_L$ and $\beta_V$ reveal the relationship between line lengths, voltage ratings, and outage rates. 
Figure \ref{fig:pdfbetas} shows the posterior distributions of $\beta_L$ and $\beta_V$ and their correlation.
The means of $\beta_L$ and $\beta_V$ are both 0.1. So the logarithm of the outage rate has a weakly positive correlation with transformed line length and transformed voltage rating. 
$\beta_L$ and $\beta_V$ have a very weak correlation, which is reasonable as $\bm x_L$ and $\bm x_V$ have a very weak correlation.

We use weakly informative priors in the Bayesian model.  If we had access to previous studies in the region, or outage rates for other similar regions then these could be used to refine the priors. In this case we would expect the uncertainty in the outage rate estimates to be reduced.

We also test somewhat stronger informative priors to check the sensitivity of the results to the prior assumptions. We reduce the standard deviation of the prior distributions of $m, \beta_L, \beta_V$ from  5 to 1 and redo the calculations. We compare the posterior mean and standard deviation of outage rates $\bm \lambda$ calculated using the different priors, and find there is not much difference. 

\subsection{Comparing the standard deviations of Bayesian and conventional estimates}

The Bayesian method produces a distribution of the outage rate, and it is straightforward to compute the standard deviation of this distribution.
The conventional method estimates the outage rate with the sample mean. The standard deviation of the sample mean can be estimated as
$s / \sqrt{n}$, where $s$ is the sample standard deviation, and $n$ is the sample size.

\begin{figure}[!ht]
\centering
\includegraphics[width=0.8\columnwidth]{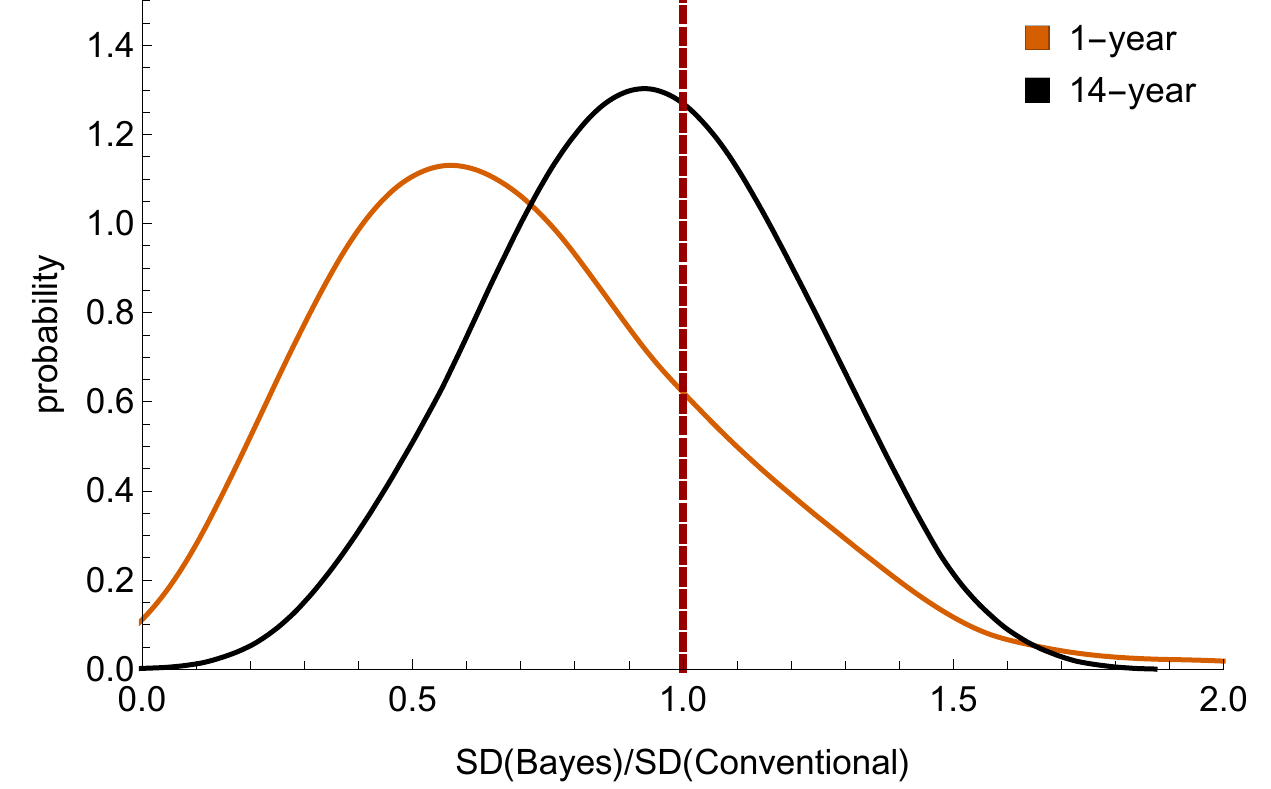}
\caption{Smoothed histograms of the ratios of standard deviations of Bayesian estimator and conventional estimator using 1 year and 14 years of data respectively. The ratio is  SD(Bayesian)/SD(conventional).}
\label{fig:cmpsdreal}
\end{figure}

Figure \ref{fig:cmpsdreal} shows the ratio of the standard deviations of the Bayesian and conventional estimators. 
It shows that the standard deviation of the Bayesian estimator is typically smaller than the conventional estimator, especially when the data is limited to one year. The median ratio of standard deviations is 0.66 for one year of data, while the median ratio is 0.93 for 14 years of data. 
Thus the Bayesian estimator typically achieves a lower standard deviation than the conventional one for limited data.
Another way to present this finding is that
given the same acceptable precision, the Bayesian method requires fewer data. Since the standard deviation is proportional to the square root of sample size,  the Bayes estimator using one year of data achieves the same standard deviation as the conventional estimator using  2.30 years of data ($1/(0.66^2)= 2.30$). 
Similarly, the Bayesian estimator using 14 years of data achieves the same standard deviation as the conventional estimator using 16.2 years of data  ($14/(0.93^2)= 16.2)$.


\subsection{Performance on rarely outaged lines}



One advantage of the Bayesian method is that it provides a principled way of making line outage rates with no observed outages. 
The conventional estimate of outage rate is zero if a line has no outage in a year. 
However, it is more reasonable that the underlying outage rate of this line is a small value. 

Table \ref{tbl:bayesest} calculates 4 line outage rates with the data available after the 1st year, after the 7th year, and after the 14th year. In Table \ref{tbl:bayesest}, line 29 has no outage except in the 9th and 10th year. 
The Bayesian estimate of the outage rate of line 29 for the 1st year is 0.32, which is informed by correlations with other lines. By the 7th year, more years with no outages have been observed, so that the estimated outage rate decreases to 0.17. 
Line 29 outages several times in the 9th and 10th years, so its estimated rate over 14 years increases.
There are also many zeros for lines 11 and 2, but the two outage rates vary differently as the distribution of zeros has different patterns. 
Most counts for line 11 are zeros, and single outages appear every several years. 
So we believe that the outage rate is roughly constant and small, which is captured by the Bayesian estimator. At the beginning, line 2 had several outages, and then it stops having outages. So this line has a decreasing outage rate. Line 8 is an example of a line with a high and increasing outage rate.


\subsection{Validation of the Bayesian hierarchical model}
Section \ref{sec:kernels} fits a linear regression model to the data, and Figure \ref{fig:residual} shows that the assumptions for this regression model hold. This validates that the form of the Bayesian hierarchical model (particularly for (\ref{equ:bayesmodel1}), (\ref{equ:bayesmodel2})) is reasonable. 

As we do not know the true outage rates using real data, we generate synthetic data to further validate the  Bayesian model in Section \ref{sec:synthetic}. 
That is, assuming that the real outage data follow the model detailed in \ref{sec:kernels}, we test that the Bayesian model accurately estimates the outage rates. As we have checked in Figure \ref{fig:residual} that the model in \ref{sec:kernels} is a good fit to the real outage data, this is a reasonable method for validating the model when we do not have the true outage rates.

\section{Test Bayesian estimates on synthetic data}
\label{sec:synthetic}
We build a generative model for synthetic datasets of arbitrary size, so the data are not limited in size, and the ground truth values are known. Then we test the Bayesian hierarchical model and the conventional estimates on the synthetic data. 
It turns out that the Bayesian hierarchical model predicts the outage rates well, and the Bayesian estimates compare favorably with the conventional method. 


\subsection{The generative model for the synthetic data}
In Section \ref{sec:kernels}, we fit a linear regression model with correlated lines. Based on this model, we generate outage counts according to the following model: 
\begin{align}
N_i &\sim {\rm Poisson}(\lambda_i G) \label{equ:generateModel1}\\
G &\sim {\rm Gamma}(a,a) \label{equ:generateModel2}\\
\ln \bm \lambda &\sim \mathcal{N}(m \mathbf{1} + \beta_L \bm x_L + \beta_V \bm x_V, \bm{\Sigma}) \label{equ:generateModel3}
\end{align}
%
The parameters in (\ref{equ:generateModel1}--\ref{equ:generateModel3}) are assigned values according to the linear regression model with correlated lines. That is, $m = -1.5$, $\beta_L = 0.13, \beta_V = 0.12$, and $\bm \Sigma = 0.52 \bm \Sigma_1 + 0.48\bm \Sigma_2$, which models the line dependencies.

Once we draw a sample from (\ref{equ:generateModel3}), the failure rate is known and fixed. So the variation of outage counts comes from the Poisson and Gamma distributions.
In particular, using $EG=1$, we derive from (\ref{equ:generateModel1}), (\ref{equ:generateModel2}) that the mean of $N_i$ is the same as only using a Poisson distribution and that $a$ controls the overdispersion:
\begin{align}
{\rm E}N_i & = {\rm E}[{\rm E}[N_i|G]]=\lambda_i  \\
{\rm Var}N_i &= {\rm E}[{\rm Var}[N_i | G]] + {\rm Var}[{\rm E}[N_i | G]]= \lambda_i + \lambda_i^2/a
\end{align}
The value of $a$ is chosen so that the variance of the model matches the empirical variance calculated from the data. In particular, 
we find the quadratic that best fits the relationship between the empirical variance and mean to be $\sigma^2 = 0.14 + 0.54 \mu + 0.53 \mu^2$ (where $\sigma^2$ is the variance, $\mu$ is the mean). 
Since the coefficients of $\mu$ and $\mu^2$ are close, we choose $a=1$.

We generate three datasets with different sizes so that we have the equivalents of 1-year, 5-year, and 100-year data: \\1) draw a sample of $\ln \bm \lambda$ from the multivariate normal distribution (\ref{equ:generateModel3}); 2) draw a sample of $G$ from the Gamma distribution (\ref{equ:generateModel2}); 3) draw samples of $N_i$ from the Poisson distribution (\ref{equ:generateModel1}) $n$~times ($n \in \{1,5,100\}$). 
Thus, we obtain $n$ annual outage counts for each line, and we know the true values of the outage rates $\bm \lambda$.

\subsection{Comparing to the conventional estimates}
The conventional estimates of outage rates are average outage counts per year. The conventional estimates and their standard deviations are obtained using Monte Carlo simulation:  draw $B=1000$ samples according to model (\ref{equ:generateModel3}), calculate the average count of each sample, and then calculate the standard deviation of the estimates.

We apply the Bayesian hierarchical model to synthetic datasets using MCMC with the same configuration as in Section \ref{sec:model}, and use the mean of the posterior distribution as a point estimate. 

\subsubsection{Errors of point estimates}

\begin{figure}[!ht]
	\centering
	\includegraphics[width=0.8\columnwidth]{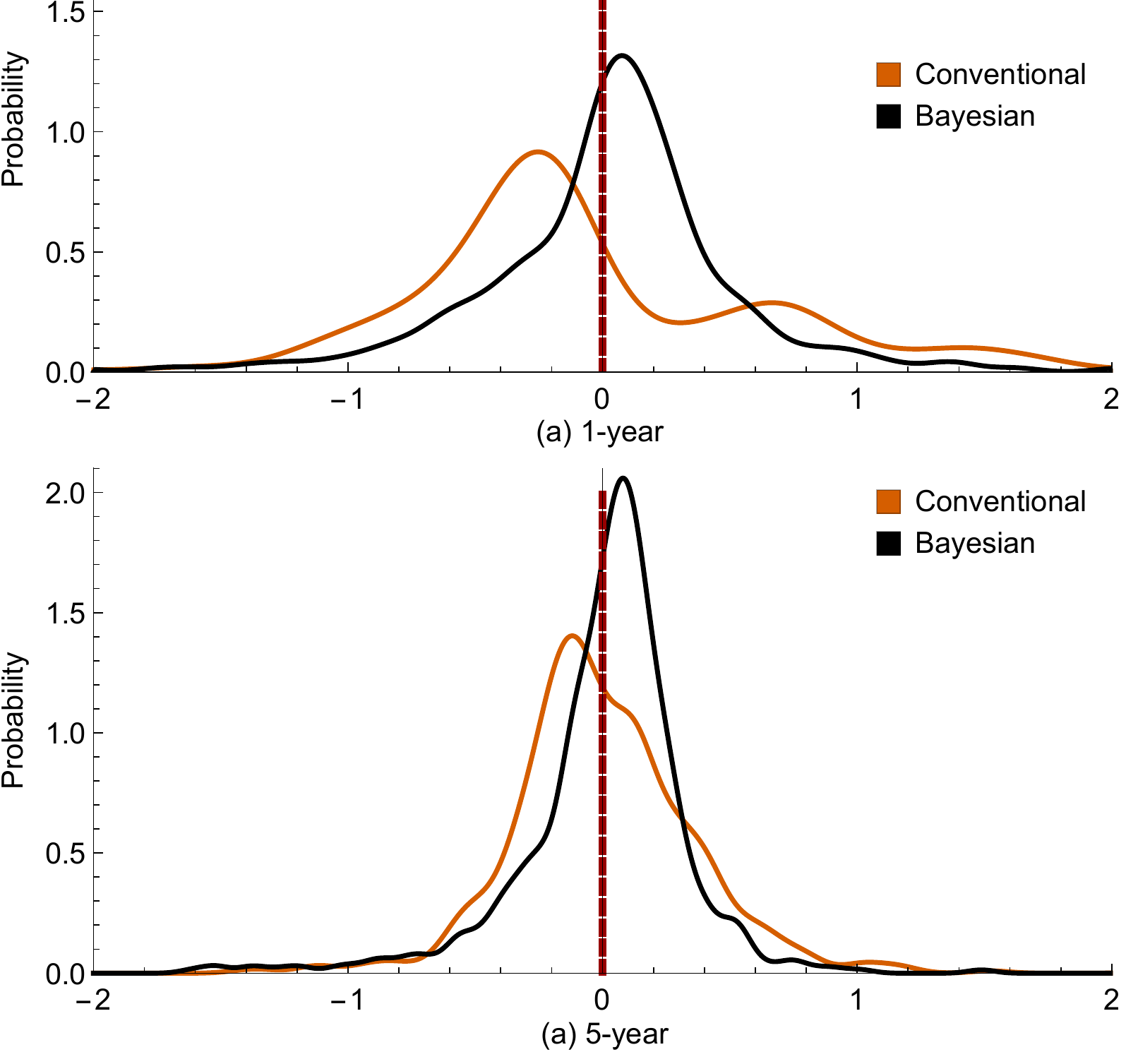}
	\caption{The distribution of point estimation errors of Bayes estimates (posterior mean) and conventional estimates using 1-year and 5-year data.}
	\label{fig:cmperrorfake}
\end{figure}

Figure \ref{fig:cmperrorfake} shows the distribution of errors of the Bayesian estimates and the conventional estimates (these estimates coincide for the 100-year data, so that the plot is not shown). In general, the less the data, the wider the histogram. The error of the conventional estimates has two modes, and the probability of error near zero is lower for \text{1-year} data. As the data size increases, the two modes merge into one. Moreover, for 1-year data, the standard deviation of the error is 0.6 for Bayesian estimates and 0.9 for conventional estimates; for \text{5-year} data, the standard deviation is 0.3 for Bayesian estimates and 0.4 for conventional estimates. Therefore, the Bayesian estimates have a high chance of obtaining more accurate point estimates, especially when data is limited. 

On the other hand, there is not much difference in the bias. Specifically, the bias is $-0.007$ for Bayesian estimates and $-0.004$ for conventional estimates using 1-year data, and the bias is $0.003$ for both Bayesian estimates and conventional estimates using 5-year data. 

\subsubsection{Standard deviation}
\begin{figure}[!ht]
	\centering
	\includegraphics[width=0.8\columnwidth]{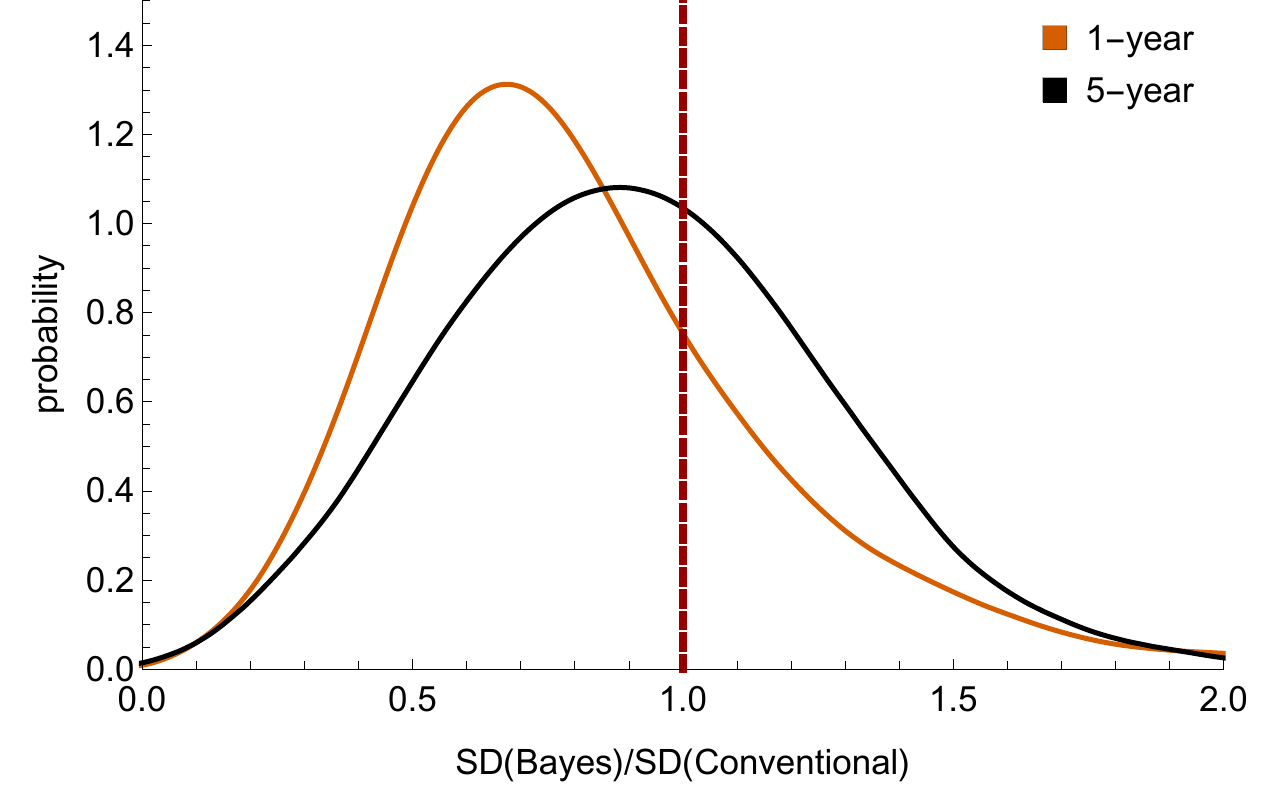}
	\caption{Distributions of the ratios of standard deviation of Bayes estimator and conventional estimator. The ratio is  SD(Bayes)/SD(conventional).}
	\label{fig:compareSDBayesMle}
\end{figure}

Figure \ref{fig:compareSDBayesMle} shows the distribution of the ratio of the standard deviation of the Bayesian estimator to that of the conventional estimator. The Bayesian estimator has a lower standard deviation when the data set is smaller. Specifically, the median of the ratio is 0.74 for 1-year data, 0.90 for 5-year data, and 0.99 for 100-year data. 
 
\subsubsection{Interval estimates}
\begin{figure}[!ht]
	\centering
	\includegraphics[width=\columnwidth]{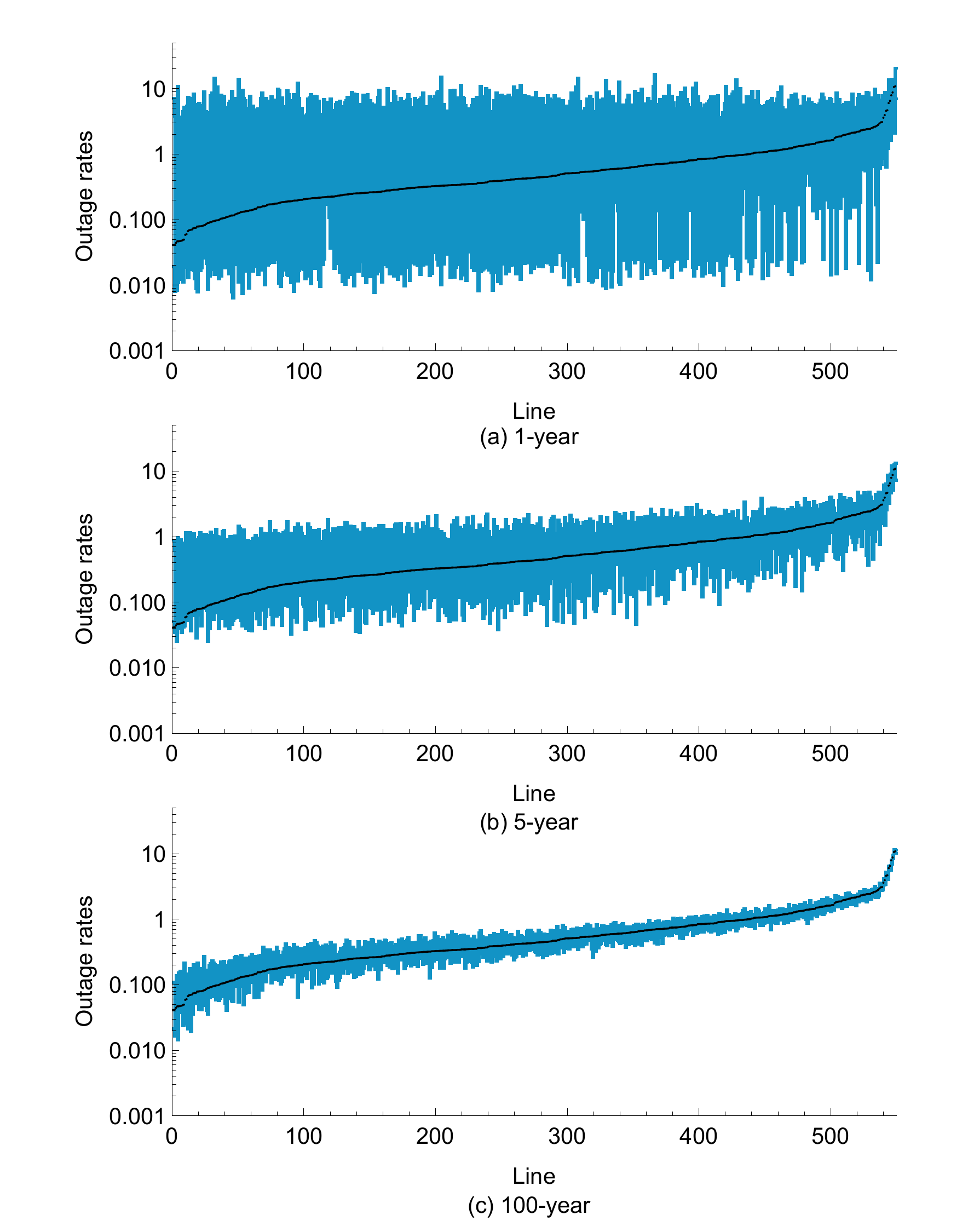}
	\caption{95\% credible intervals of Bayesian estimates using 1-year, 5-year and 100-year data. Lines are ordered by outage rates (black dots).}
	\label{fig:BayesMleCI}
\end{figure}

Figure \ref{fig:BayesMleCI} shows 95\% credible intervals of the Bayes estimator  using 1-year, 5-year, and 100-year data respectively. 
As the size of the dataset increases, we gain more information, and the width of the credible intervals decreases.
Figure \ref{fig:BayesMleCI} also shows the true values of the outage rates as black dots. As expected with a 95\% credible interval, approximately 5\% of the true values lie outside the credible interval. 
The Bayesian point estimates (not indicated  in Figure \ref{fig:BayesMleCI}) lie in the center of the credible intervals and tend to be larger than the true values for low outage rates and smaller than the true values for high outage rates. This can be explained as the shrinkage towards the mean expected with Bayesian methods; see \cite[Sec. 1.5]{carlin08BMDA}.

\section{Discussion and Conclusion}
\label{sec:conclusion}
We use a Bayesian hierarchical model to improve the estimation of annual outage rates for individual transmission lines. This Bayesian method incorporates several types of dependencies between lines and is applied to real outage data and tested with synthetic data. Particularly for the shorter observation periods with the lower outage counts,  the Bayesian estimates perform better than the conventional estimates that simply divide the number of outages by the observation time: estimates of the individual line outage rates are more accurate, and the uncertainty of the estimates is reduced. 

Our Bayesian hierarchical model offers an improvement over the conventional estimates for two reasons. Firstly, the Bayesian method can appropriately capture our prior knowledge of the parameter uncertainties with prior distributions. 
Secondly, because the model is hierarchical and models the dependence between lines, information about multiple partial commonalities can be appropriately shared across similar lines. These reasons imply that estimates can be improved for lines with no (or a small number of) outages. 

Geographically close and neighboring lines experience similar weather conditions, may have a similar design, and share some physical and engineering interactions through the network.
We model these line dependencies as a covariance matrix in the Bayesian hierarchical model. The covariance matrix is the weighted sum of two kernels that represent  geographic district commonalities and network line proximity, respectively. 
The Bayesian model learns the weights of the two kernels from the outage data. 
Our modeling of these dependencies can be realized from a single utility outage dataset that is routinely collected, since the line district is recorded in the dataset, and the network can be readily deduced from the dataset \cite{DobsonPS16}.

Previous work has often assumed that transmission line outage rates are proportional to line length \cite{IesmantasEPSR19} or grouped together lines of the same area \cite{ZhouPS06,LiIBM10,DokicHICSS19}.
We model these dependencies by linear factors in the outage rate, and the Bayesian model learns the weights for these factors. The results for our data are that individual line outage rates are only partially correlated with the line length or the voltage rating. Therefore, it is more reasonable to consider the outage rate for a whole line instead of the rate per mile. 


The Bayesian method estimates the distribution of individual line outage rates. This is an advantage compared to methods that return point estimates, as a complete picture of the uncertainty around estimates is needed to make robust decisions about risk and maintenance.
For example, if a line has a high point estimate outage rate that is very uncertain, it may be beneficial to wait to gather more information. 
If desired, any point or interval estimates can be easily obtained from the distribution, depending on the desired application of the outage rates. 
The quantification of the uncertainty of estimates is useful when the outage rates are used in other models and simulations.
For example, a Monte Carlo simulation of transmission reliability can easily be modified to sample from the outage rate distribution to better capture the uncertainty in the estimated reliability.

We focus on overall line outage rates without considering different outage causes in this paper. However, the proposed Bayesian method can naturally be extended to investigate line outage rates for specific causes. 

\looseness=-1
When data is limited, which is generally true for power system outage data, Bayesian estimates have smaller uncertainty than conventional estimates. 
Equivalently, with a specific acceptable standard deviation, the proposed Bayesian method needs less data than the conventional method. Thus, utilities can monitor individual line outage rates with fewer years of recording outages. There is a potential to more quickly identify lines with increasing outage rates and aging problems so that maintenance can be scheduled. For example, if utilities need two years of data using the conventional method to estimate line outage rates with a given uncertainty, they typically only need one year of data using the proposed Bayesian method to obtain an outage rate estimate that meets the same uncertainty requirement.


\appendices




\section{Model diagnostics}
\label{app2}
\begin{figure}[!ht]
	\centering
	\includegraphics[width=\columnwidth]{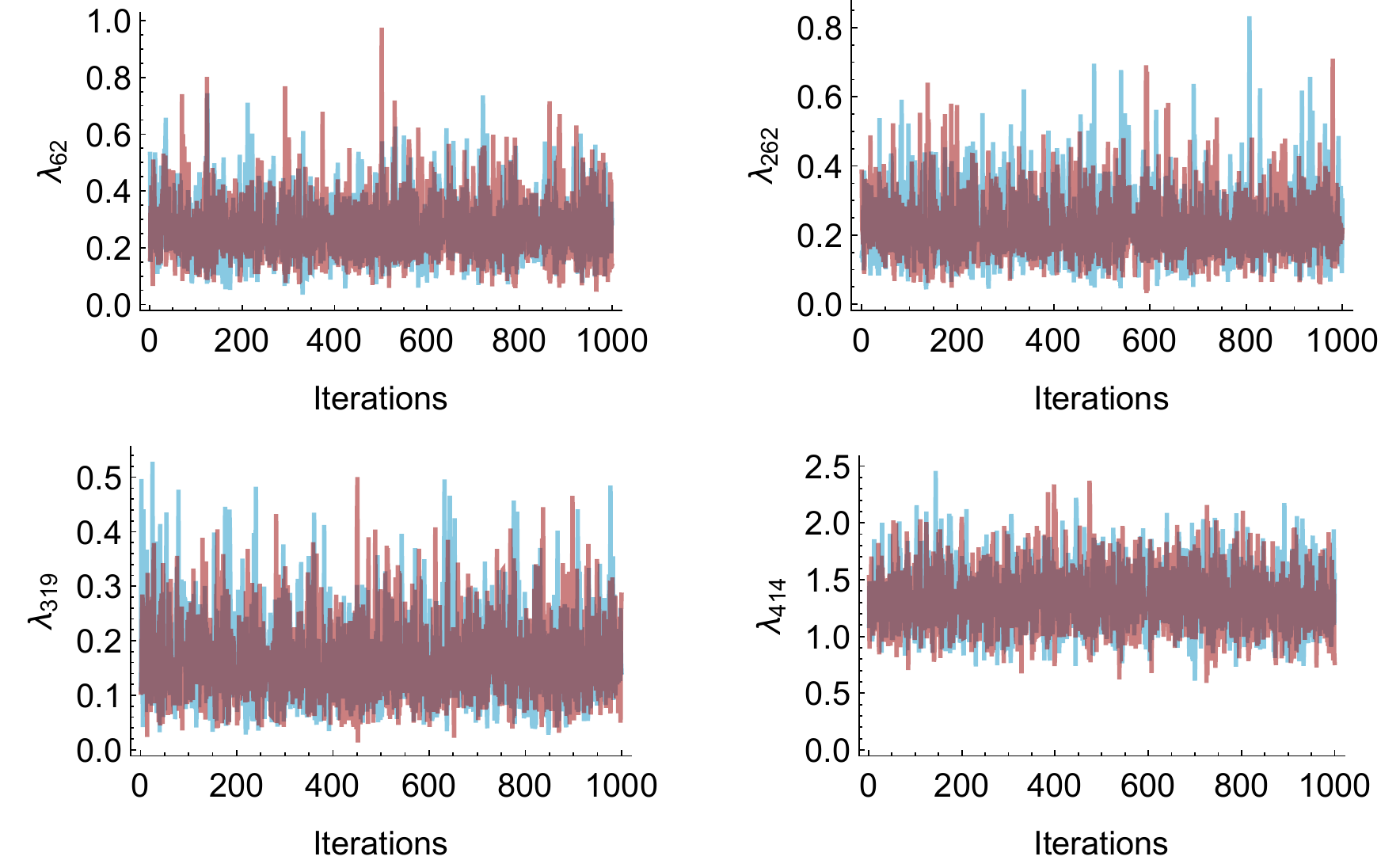}
	\caption{Trace plots of two chains of four randomly selected parameters .}
	\label{fig:traceplot}
\end{figure}

\begin{figure}[!ht]
	\centering
	\includegraphics[width=\columnwidth]{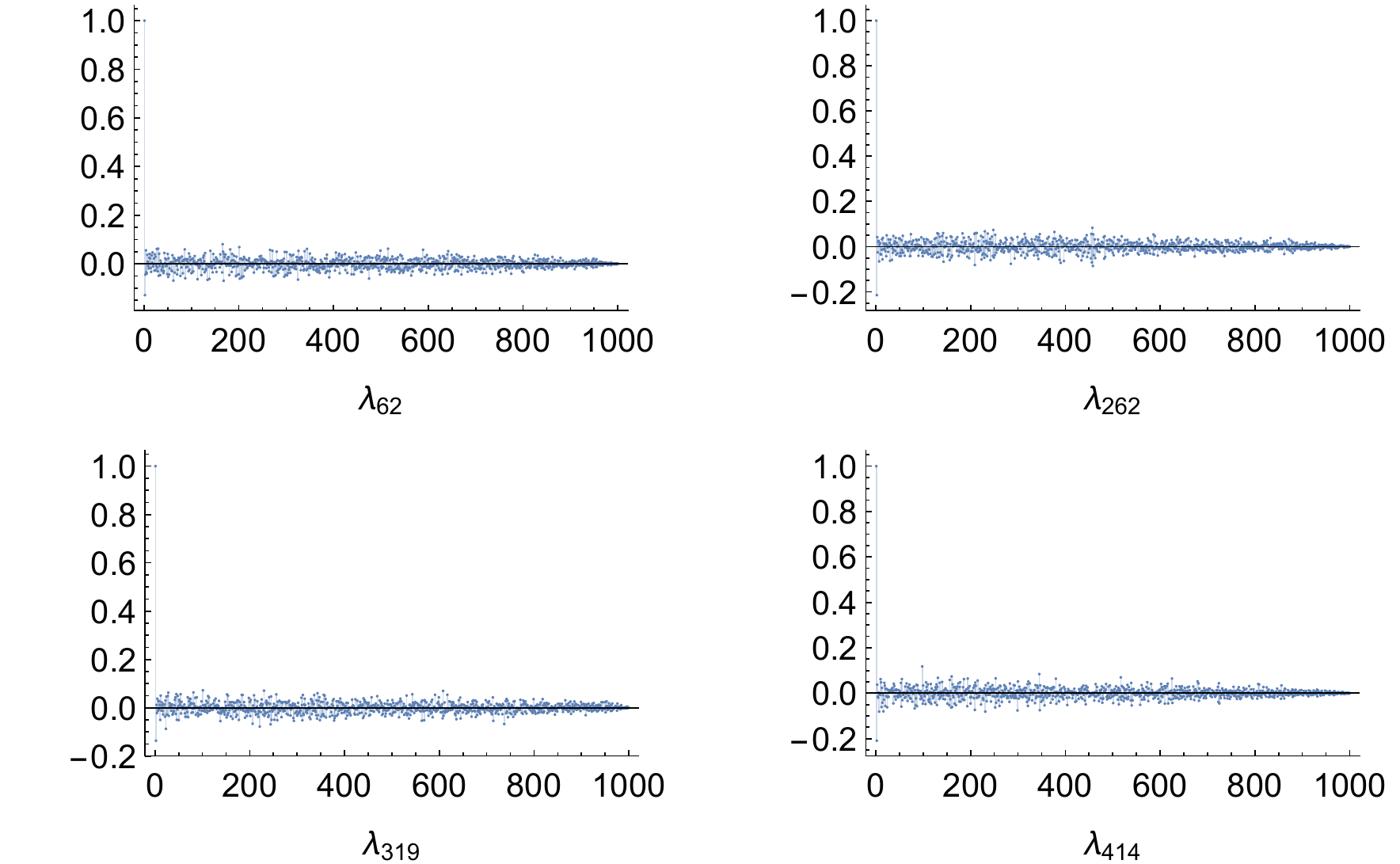}
	\caption{Autocorrelation function plots of four randomly selected parameters.}
	\label{fig:ACFplot}
\end{figure}

\looseness=-1
There is no evidence of nonconvergence \cite[Sec. 3.4]{carlin08BMDA} \cite[Sec. 11.4]{GelmanBDA13}. The potential-scale-reduction statistics $R^{hat}<1.06$, and the ratio of number of effective samples $N_{eff}/N>0.004$.
As there are 1657 parameters, it is not practical to show all the trace and autocorrelation plots. Here we randomly select four parameters to show the trace plots (Figure \ref{fig:traceplot}) and autocorrelation function plot (Figure  \ref{fig:ACFplot}). The two chains have mixed, and the autocorrelation decreases quickly and tends to zero. 


\bibliographystyle{IEEEtran}
\bibliography{IEEEabrv,reference}

\end{document}